\documentclass[11pt,journal,comsoc]{IEEEtran}

\usepackage[T1]{fontenc}
\usepackage{amsmath}
\usepackage{cite}
\usepackage{amsmath,amssymb,amsfonts}
\usepackage{algorithmic}
\usepackage{graphicx}
\graphicspath{{/fig/}}
\DeclareGraphicsExtensions{.pdf,.jpeg,.png}
\usepackage{tabularx}
\usepackage{textcomp}
\usepackage{xcolor}
\usepackage[cmintegrals]{newtxmath}

\def\BibTeX{{\rm B\kern-.05em{\sc i\kern-.025em b}\kern-.08em T\kern-.1667em\lower.7ex\hbox{E}\kern-.125emX}}
\usepackage{subcaption}

\makeatletter
\def\ps@IEEEtitlepagestyle{%
  \def\@oddfoot{\mycopyrightnotice}%
  \def\@oddhead{\hbox{}\@IEEEheaderstyle\leftmark\hfil\thepage}\relax
  \def\@evenhead{\@IEEEheaderstyle\thepage\hfil\leftmark\hbox{}}\relax
  \def\@evenfoot{}%
}
\def\mycopyrightnotice{%
  \begin{minipage}{\textwidth}
  \centering \scriptsize
  Copyright~\copyright~2022 IEEE. Personal use of this material is permitted. Permission from IEEE must be obtained for all other uses, in any current or future media, including reprinting/republishing this material for advertising or promotional purposes, creating new collective works, for resale or redistribution to servers or lists, or reuse of any copyrighted component of this work in other works by sending a request to pubs-permissions@ieee.org.
  \end{minipage}
}
\makeatother

\begin{document}

\title{Autonomous UAV Base Stations for Next Generation Wireless Networks: A Deep Learning Approach}
%\title{Autonomous UAV Base Stations for 6G Wireless Networks: A Deep Learning Approach}
\author{Ali~Murat~Demirtas,~\IEEEmembership{Member,~IEEE,} Mehmet~Saygin~Seyfioglu,~\IEEEmembership{Student~Member,~IEEE,} Irem~Bor-Yaliniz,~\IEEEmembership{Senior~Member,~IEEE,} Bulent~Tavli,~\IEEEmembership{Senior~Member,~IEEE,} and Halim~Yanikomeroglu,~\IEEEmembership{Fellow,~IEEE}
\thanks{A. M. Demirtas and B. Tavli are with the Department of Electrical and Electronics Engineering, TOBB University of Economics and Technology, Ankara, Turkey (e-mail: ademirtas@etu.edu.tr; btavli@etu.edu.tr).}
\thanks{M. S. Seyfioglu is with the Department of Electrical and Computer Engineering, University of Washington, Seattle, USA (e-mail: msaygin@cs.washington.edu).}
\thanks{I. Bor-Yaliniz and H. Yanikomeroglu are with the Department of Systems and Computer Engineering, Carleton
University, Ottawa, Canada (e-mail: irembor@sce.carleton.ca; halim@sce.carleton.ca).}}

%\markboth{IEEE Vehicular Technology Magazine}%
%{Demirtas \MakeLowercase{\textit{et al.}}: Autonomous UAV-Mounted BSs for 6G Wireless Networks: A DL Approach}

\maketitle

\begin{abstract}
\label{section:abstract}To address the ever-growing connectivity demand in communications, the adoption of ingenious solutions, such as utilization of Unmanned Aerial Vehicles (UAVs) as
mobile Base Stations (BSs), is imperative. In general, the location of a UAV Base Station (UAV-BS) is determined by optimization algorithms, which have high computationally complexities and are hard to run on UAVs due to energy consumption and time constraints. In this paper, we show that a Convolutional Neural Network (CNN) model can be trained to infer the location of a UAV-BS in real time. To this end, we create a framework to determine the UAV locations considering the deployment of Mobile Users (MUs) to generate labels by using the data obtained from an optimization algorithm. Performance evaluations reveal that once the CNN model is trained with the given labels and locations of MUs, the proposed approach is, indeed, capable of approximating the results given by the adopted optimization algorithm with high fidelity, outperforming Reinforcement Learning (RL)-based approaches.  We also explore future research challenges and highlight key issues.

\begin{IEEEkeywords}
Next generation wireless networks, UAV, deep learning, CNN, reinforcement learning.
\end{IEEEkeywords}
\end{abstract}

\section{Introduction}
\label{section:introduction}

%Ubiquitous and high data rate connectivity has become an indispensable aspect of human life, which has only been exacerbated by the current COVID pandemic. Indeed, the Internet-of-Things (IoT), which is a key constituent of many contemporary paradigms, such as Industry 4.0, smart cities, and autonomous connected vehicles, is becoming more and more ubiquitous. Therefore, the ever growing need for higher data rates and better connectivity across highly challenging scenarios and environments necessitates the adoption of extraordinary technological solutions.

Ubiquitous and high data rate connectivity has become an indispensable aspect of human life. Indeed, many contemporary paradigms, such as Industry 4.0, smart cities, and autonomous connected vehicles, is becoming more and more ubiquitous. Therefore, the ever growing need for higher data rates and better connectivity across highly challenging scenarios and environments necessitates the adoption of extraordinary technological solutions.

Recently, Self-Evolving Networks (SENs)~\cite{darwish2020vision} have been proposed to facilitate autonomous network management to reduce human intervention as much as possible by considering the agility and Machine Learning (ML)-based management aspects of 
%6G Wireless Networks (6GWNs). 
Next Generation Wireless Networks (NGWNs).
Integration of terrestrial, aerial, and satellite networks to form a vertical heterogeneous network, which can be considered as an evolving architecture, has a potential to incorporate significant agility into the NGWN infrastructure. 
%However, resource allocation and network management under such highly dynamic conditions arise as challenging research problems. 
However, network management under such highly dynamic conditions arises as a challenging research problem. 
\begin{figure*}[hbt]
	\centering
    \vspace{-0.5cm}
    \begin{center}
        %\shorthandoff{=}
        \includegraphics[scale=0.55]{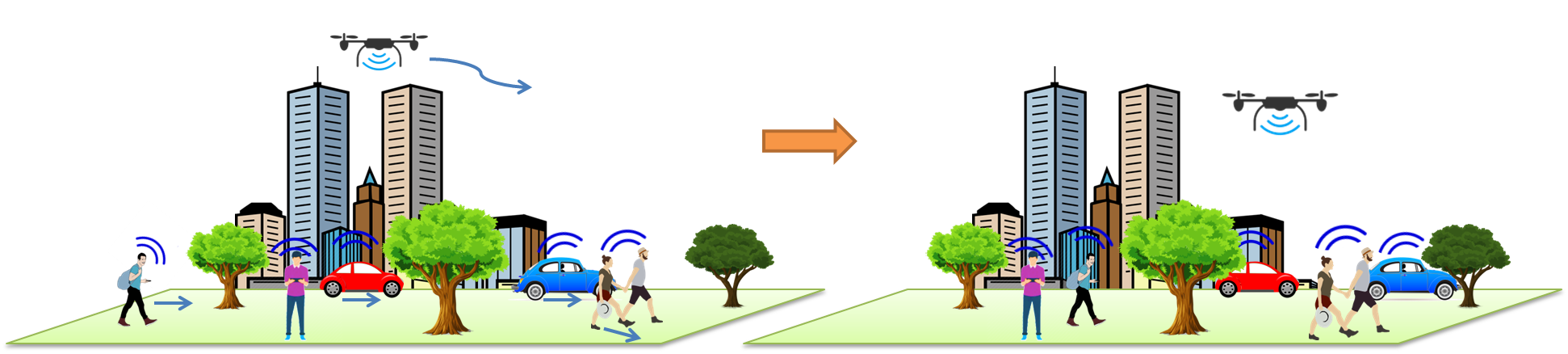}
        \caption{Illustration of a representative communications scenario.}
        \label{figure:model-overview}
        %\shorthandon{=}
    \end{center}
    \vspace{-0.7cm}
\end{figure*}

Within the framework of SENs, Unmanned Aerial Vehicle Base Stations (UAV-BSs) are designated as highly dynamic infrastructure enablers and their potential advantages are enormous. The inherent availability of Line-of-Sight (LoS) links between a UAV-BS and Mobile Users (MUs) can significantly reduce the path-loss, which in turn leads to greater coverage area and/or better channel capacity. Furthermore, communications technologies such as millimeter wave and free space optics can benefit from the availability of LoS links. Unprecedented mobility associated with UAV-BSs enable the network as a whole to behave like a highly dynamic entity (see Fig.~\ref{figure:model-overview}), which is one of the main pillars of SENs. Hence, it is possible to restructure the network on an unprecedented scale and level that it can be possible to create an extremely flexible and reliable mobile communication infrastructure. For example, UAV-BSs can be utilized to augment communications infrastructure in locations (especially in metropolitan areas) 
%where short-term demand is hugely inflated such as entertainment gatherings, social meetings, or sport events. 
where there may be significant short-term peaks in demand, such as during large entertainment and sporting events.
In fact, the Third Generation Partnership Project (3GPP) has already defined an on-board radio access node (UxNB) in recognition of the great potential of UAV-BSs (3GPP Release 17). Indeed, the literature on the efficient utilization of UAVs as an integral part of NGWNs is rich. Applications of UAVs in NGWNs are envisioned to revolutionize all aspects of telecommunications technology profoundly. However, considering the fact that utilization of UAVs in telecommunications is still in its infancy phase, it appears that extensive further research is required to achieve technological maturity for UAV-BSs. Nevertheless, to harvest the potential advantages of UAV-BSs, 
%many technological challenges, such as adaptation to dynamic channel characteristics and user mobility, security, robustness, and energy efficiency should be addressed. 
many technical challenges still need to be addressed, such as adaptability to dynamic channel characteristics, user mobility, security, robustness, and energy efficiency.

%Due to three dimensional and high mobility and, potentially, large service area, 
Due to the three-dimensional space, high mobility, and large service area, the channel between a UAV-BS and MU can vary significantly. 
%Hence, optimizing an instance 
For this reason, optimizing a single instance of the communication scenario would 
%be a too conservative and an unrealistic approach. 
be a too conservative and an unrealistic approach.
%Therefore, an 
Instead, an autonomous UAV-BS should be able to adapt itself according to the dynamic characteristics of the communication medium and user mobility by incorporating 
%the temporal and spatial characteristics 
the relevant temporal and spatial variables into the decision-making process.

While determining the trajectory of a UAV-BS (and in any real-life problem solving approach) it would be unwise to assume an ideal environment. For it is highly unlikely that the channel estimation 
%is precise and mobility prediction is free of errors. 
would be precise and mobility prediction error-free.
Furthermore, 
%the existence of 
man-made hindrances such as inaccurate information flow or adversarial behaviors can, potentially, render the decision making process more challenging. Therefore, resilience against errors and 
%incorrect information 
threats is a necessary requirement in UAV-BS placement optimization.

The limited energy resources of UAVs are, arguably, among the most challenging problems that hinder their utilization as UAV-BSs. To get the maximum benefit out of the limited airborne duration of a UAV-BS, optimization of its trajectory, considering the requirements of the MUs it is meant to be serving, is of utmost importance. 

Addressing the aforementioned problems is extremely challenging (if not impossible) by employing conventional trajectory optimization approaches (e.g., Mixed Integer Non-Liner Programming -- MINLP). Yet, a Deep Learning (DL)-based trajectory optimization approach has the potential to cope with such challenges~\cite{Goodfellow16}. DL is an emerging ML approach that is extremely successful at extrapolating hidden non-linear models between the input and output layers using huge and cumbersome data. Despite the rather short history of the applications of DL in communications thus far, the performance attained by DL-based solutions are promising~\cite{OSheaH17}. 

In this study, we propose a DL-based UAV-BS trajectory optimization approach and evaluate its 
%performance in comparison to Reinforcement Learning (RL)-based approaches. 
performance against Reinforcement Learning (RL)-based approaches.
%The proposed approach is computationally efficient (once the DL is trained), it can cope with the user mobility, and it is both scalable and robust.
Once the DL is trained, our proposed approach is shown to be computationally efficient and capable of coping with user mobility, while also being scalable and robust.

The rest of the paper is organized as follows. In Section~\ref{section:reference_model}, we introduce a generic optimization model for the UAV-BS placement problem. Section~\ref{section:approaches} explores the approaches for the solution of the optimization model. Section~\ref{section:learning_methods}, concisely, overviews the building blocks of our DL model. A use case scenario employing the DL model to solve the optimum coverage problem of a UAV-BS is presented and analyzed in Section~\ref{section:scenario}. In Section~\ref{section:research_challenges}, potential future research areas are discussed. Section~\ref{section:conclusion} concludes the paper.

\section{Reference Model}
\label{section:reference_model}

%Optimization of the operation of a UAV-BS 
Optimizing the operations of a UAV-BS is a challenging, multi-faceted problem. Indeed, the problem has many aspects such as coverage, capacity, Quality-of-Service (QoS), caching, energy dissipation, and dynamic channel and MU characteristics. A generic reference model for the optimization of UAV-BS operations can be expressed as follows:
\begin{equation}
\begin{gathered}
\max \sum_{k \in \mathbb{U}, t \in \mathbb{T}}^{} w_{1} u_{k,t} + w_{2}  M(z_{k,t},u_{k,t}) - w_{3}  E(V_{t},a_{t}). \\
\end{gathered}
\label{eq:solve} 
\end{equation}
The model has three objectives, where the parameters $w_{1}$, $w_{2}$, and $w_{3}$ denote the relative importance of each objective. \textit{Coverage} (i.e., serving as many MUs as possible), arguably, is the foremost objective in our problem, which can be achieved by positioning the UAV-BS, at a given time instant, at a position (defined by the vector $\textbf{d}_{t}$) that results in the maximum number of users experiencing pathloss values lower than a predetermined threshold ($\gamma$)~\cite{YalinizEY16}. In the model, subscript $t$ signifies the temporal characteristics of the problem to account for the \textit{dynamics of the channel and the characteristics of MUs} as well as the \textit{mobility of the UAV-BS}. The time horizon and the set of MUs are denoted by $\mathbb{T}$ and $\mathbb{U}$, respectively. The variable $u_{k,t}$ is set to one if the MU-$k$ is served at time instant $t$. The bandwidth available to a UAV-BS is limited, therefore, the aggregated \textit{capacity} allocated to MUs at each time instant $t$ is upper bounded by the total available backhaul capacity~\cite{KalantariSYY17}.

\textit{QoS} provided to the MUs is an important consideration~\cite{AlzenadEY18}. The MUs' QoS requirements can be heterogeneous (e.g., QoS requirements for email traffic, phone calls or video exhibit significant variations) and can change in time (i.e., dynamic). The QoS provided to each MU~$k$ depends on $\textbf{d}_{t}$, and is constrained by the QoS requirement of MU~$k$ at time instant $t$. The QoS criteria encompasses many metrics, such as bandwidth and delay.

It is possible 
%that multiple MUs demand 
for multiple MUs to demand the same 
%content, hence, \textit{caching} popular content 
content, and for this reason \textit{caching} popular content
is beneficial in reducing the bandwidth usage as well as delay~\cite{ChenMSYDH17}. However, caching a high volume of content is not possible with a resource-constrained UAV-BS. Therefore, the utility function for caching, $M(.)$, which depends on MUs to be served ($u_{k,t}$) and the content demand of the MUs ($z_{k,t}$), should be maximized.

The \textit{energy dissipation} characteristics of the UAV-BS operation is of utmost importance since the service duration of a UAV-BS is, generally, determined by its power usage. Indeed, in certain cases the service duration of the UAV-BS is more important than the coverage or QoS. Therefore, 
%energy dissipation, $E(.)$, as a function of speed, $V_{t}$, and acceleration, $a_{t}$, of the UAV-BS is 
energy dissipation, $E(.)$, of the UAV-BS, as a function of its speed, $V_{t}$, and acceleration, $a_{t}$, is embedded as a separate term in the objective function. 

\section{Solution Approaches}
\label{section:approaches}

The solution of the reference model to optimality is highly challenging. In fact, depending on the functions utilized in the objective and constraints, the problem can turn out to be non-convex. Furthermore, the temporal dimension of the problem can hugely inflate the size of the solution space, which in turn can greatly increase the computational complexity of the solution algorithms. Solving a single time instance of the problem, which is a commonly employed approach, necessitates resolving the problem for each time instant. Solutions 
%considering only a single time instant 
that consider only a single time instance
can result in suboptimal solutions (for a finite time interval -- not an instance) which can lead to unnecessary movements of the UAV-BS. Moreover, due to errors in collecting or obtaining data (e.g., positions of MUs, pathloss values), solution approaches which cannot tolerate such non-ideal inputs can potentially end up yielding inconsistent or inefficient UAV-BS trajectories.  

To overcome such challenges efficiently, an ML-based approach can be created to estimate the optimal location of a UAV-BS. 
%RL-based approaches are proposed, in the literature, to determine UAV-BS locations~\cite{chen2019artificial,ullah2020cognition,lahmeri2021artificial}. 
In RL~\cite{tang2021survey}, there is an agent which can be found in one of the predefined states. The agent takes actions to move among different states and the actions are rewarded. For each state, the agent decides the action to maximize the total future reward and keep it in the state-action table. RL guarantees the best actions only if the number of iterations is large enough, therefore, convergence can potentially require formidably long time periods depending on the solution space size. Moreover, if RL keeps state-action pairs for all deployment scenarios, then it will be impractical for a resource constrained UAV-BS to adopt an RL-based approach due to limited memory. Deep Reinforcement Learning (DRL)~\cite{wang2020thirty} is a form of RL that estimates state-action pairs using a Deep Neural Network (DNN) to mitigate the memory requirements of huge state state-action tables. Hence, DNN-based approaches can be used to alleviate the memory requirements for RL\cite{tang2021survey,chen2019artificial,ullah2020cognition,zappone2019model,lahmeri2021artificial}. However, the training time for the establishment and convergence of a fully connected network for each state-action pair of all possible UAV-BS and user deployments can be impractically long. 
%\textbf{In summary, RL based approaches work well in unknown environments with stationary users, whereas we consider a solution for an environment  with immobile structures and dynamic MUs movement for efficient UAV-BS deployment.}
%In summary, RL-based approaches work well in unknown environments with stationary users, 
%whereas we consider a solution for an environment with dynamic MUs for efficient UAV-BS deployment.
%but in this paper we consider an environment with dynamic MUs for efficient UAV-BS deployment. 

\begin{figure*}[!h]
    %\vspace{-0.1cm}
    \includegraphics[width=\textwidth,height=5cm]{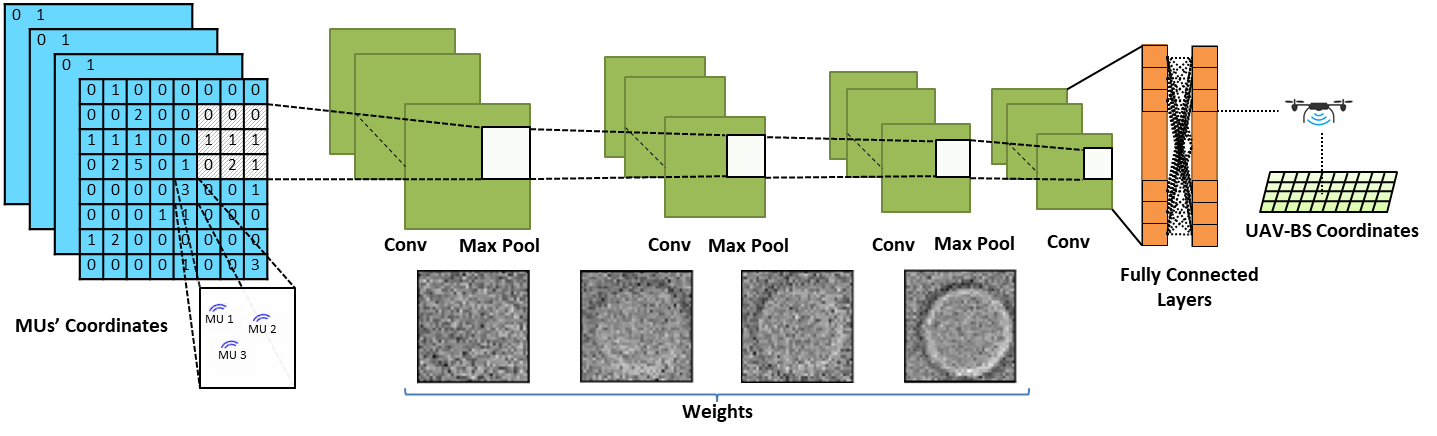}
    \caption{The CNN architecture.}
    \label{fig:architecture}
    %\vspace{-0.3cm}
\end{figure*}
 
A Convolutional Neural Network (CNN) is an Artificial Neural Network (ANN) with multiple hidden layers (more precisely, multiple convolutional layers, where the data from the previous layer is convolved with different filters). Indeed, CNN-based 
%approaches are shown to be performing well 
approaches have been shown to perform well in solving a wide range of complex problems that could not be otherwise solved efficiently. Therefore, in this study, %we opted to create a CNN-based 
we opt for a CNN-based DL approach to determine the trajectory of an autonomous UAV-BS. 

The most computationally demanding part of the proposed approach is the training stage, which is performed offline. 
%Once trained sufficiently, 
Once the model has been trained sufficiently, the solution can be obtained almost instantly. Since the CNN-based approach has a holistic view of the problem, the UAV-BS trajectory is created by considering multiple 
%time instants at once. 
time instances simultaneously. The proposed model is resilient to varying channel characteristics and movements of MUs (i.e., once the CNN is trained satisfactorily, the trajectory is computed effortlessly for any input vector without the need to rerun an optimization algorithm). Since the CNN filters the input data to eliminate noise and noise-like inputs by considering an extended period of time, it can tolerate errors and incorrect information present in the inputs to a certain extent.

\section{Convolutional Neural Networks}
\label{section:learning_methods}

DL~\cite{Goodfellow16} is an ML approach 
%that recently has experienced a resurgence due to 
that has begun to receive much attention due to the unprecedented increase of computing capabilities offered by modern GPUs and advances in algorithms. 
%DNNs build on the extensive experience of ANNs by increasing the overall size of the network using many layers of neurons. Each neuron is formed by linearly weighting multiple inputs supplied to an activation function. 
Applications of DL-based approaches for communications and optimization problems have resulted in improved performance and functionality compared to existing solutions that do not utilize DL.  

One of the most successful DL architectures is CNN~\cite{LeCunBH15}, which provides superior performance benchmarks never before achieved 
%in various problems. 
for a variety of problems.
CNNs are comprised of convolutional layers, pooling layers, and fully connected layers, and they are generally used with a softmax function to perform classification tasks. CNNs excel at learning spatially localized features by the convolution of the input data with filters in a sliding window arrangement. Initial layers learn generic/simple features, whereas deeper layers learn more abstract/sophisticated features by non-linearly combining the features learned by the preceding layers. 

An activation 
%function is generally followed after each 
function generally follows each convolution operation, which performs a non-linear mapping of the convolution output. Among the plethora of proposed activation functions, we adopt the Rectified Linear Unit (ReLU), which offers both robustness and computational efficiency. After activation, 
%max pooling operation is, generally, performed which is a downsampling operation utilized to reduce 
a max pooling operation is generally performed, which is a downsampling operation done to reduce
the complexity of activation maps that are further processed in the following/deeper layers of the network. 
%At the end, a fully connected neural network is used 
Finally, at the penultimate layer (i.e., the last convolutional block), a fully connected neural network is used to learn non-linear combinations of all the learned features.
%at the penultimate layer (i.e., the last convolutional block).

\begin{figure*}[!h]
\vspace{-0.3cm}
    \hspace*{-2.4cm}
    \includegraphics[width = 8.8in]{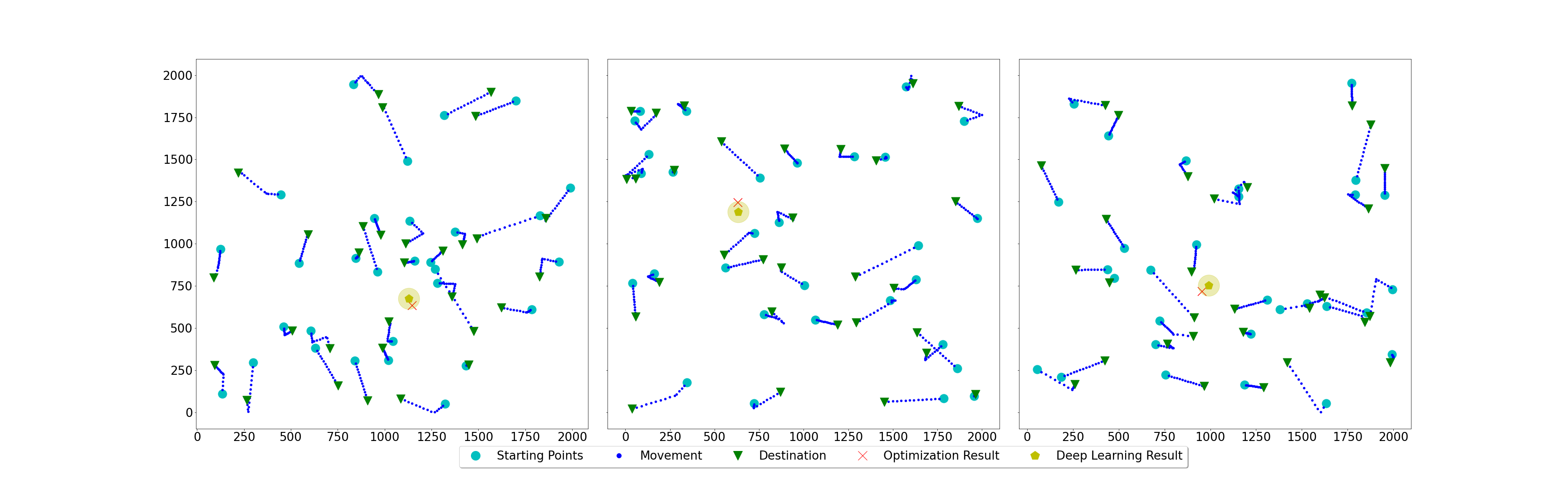}
    \caption{Locations of the UAV-BS for the optimal and CNN-based solutions.}
    \label{figure:CNN-Res}
    \vspace{-0.3cm}
\end{figure*}
We utilize a six-layer CNN architecture, which we 
%found to be producing the best results 
found produced the best results through empirical evaluations. The proposed architecture is comprised of four convolutional layers, as illustrated in Fig.~\ref{fig:architecture}. The network area is spatially organized into rectangular cells (e.g., a checkerboard structure). Furthermore, each cell encompasses five consecutive time instants in the temporal dimension. An array consisting of the numbers of MUs in each cell is fed to the model as an input. Utilizing an input, which is embedded with both temporal and spatial characteristics, has several advantages. First, the scalability of the model is not affected by the number of MUs (i.e., the size of the input does not change with the number of MUs). Second, the model is resilient to the ingress/egress of MUs to/from the area under consideration. Third, the model facilitates the use of only a subset of the MUs in the input which can be used to decrease the computational complexity. Although spatial and temporal clustering of the MUs can decrease precision, it increases the ability to generalize, which improves the robustness of the model. CNNs can exploit spatiotemporal characteristics of data more efficiently when compared to other learning approaches due to the embedding of various convolutional filters within the CNN architecture, which is one of the most important reasons for employing CNNs in our model. The output layer of the CNN model utilizes a linear activation 
%function and it is used to get the continuous outputs 
function that yields continuous outputs for UAV-BS coordinates. Adam is used for learning rate optimization and the loss function used is the mean absolute error. 
%\section{Performance Evaluation}
\section{Use Case Scenario}
\label{section:scenario}
%\textbf{ We generate a UAV-BS deployment scenario to analyze the performance of the CNN based solution. This scenario only considers coverage aspect of the reference model explained in Sec.~\ref{section:reference_model} but it is straight forward to extend it to other aspects of the model like QoS and caching.}   
We generate a UAV-BS deployment scenario to analyze the performance of the CNN-based solution. This scenario only considers coverage aspect of the reference model explained in Section~\ref{section:reference_model}, however, it is also possible to take the other aspects of the model, such as QoS and caching, into account.
To train the CNN, we create a synthetic dataset. First, we obtain the optimal locations of the UAV-BS for a range of MU deployments using reference model parameters~\cite{YalinizEY16}. Next, we use the constraints and locations of MUs as inputs and optimal locations of the UAV-BS as the output to train our CNN model. Lastly, we test the CNN model to compute the locations of the UAV-BS for new user deployments (i.e., the test cases). Moreover, we provide comparisons of the solutions obtained with the CNN approach and various RL approaches.

%We demonstrate the efficiency of the proposed CNN-based approach by using a typical urban environment communications scenario served by a UAV-BS. 30 MUs are distributed over a 2~km by 2~km area. The locations of MUs are within the radius of a predetermined center, which is found using a uniform random distribution. Each MU moves toward a randomly chosen direction with a constant speed, which is also determined randomly. For this, we employed the well-known random-waypoint mobility pattern~\cite{HyytiaV07}. A series of 15 consecutive moves of an MU is called a session. MUs restart their movements at random positions at the beginning of each session. Our scenario consists of 72,000 sessions. The rest of the communication parameters are adopted from~\cite{YalinizEY16}.
We demonstrate the efficiency of the proposed CNN-based approach by using a typical urban environment communications scenario served by a UAV-BS. 30 MUs are distributed over a 2~km by 2~km area. The locations of MUs are within the radius of a predetermined center, which is found using a uniform random distribution. Each MU moves toward a randomly chosen direction with a constant speed (i.e., random-waypoint mobility model~\cite{HyytiaV07}). A series of 15 consecutive moves of an MU is called a session. MUs restart their movements at random positions at the beginning of each session. Our scenario consists of 72,000 sessions. The rest of the communication parameters are adopted from~\cite{YalinizEY16}.

For each instant of the scenario, we compute the best UAV-BS position,
which will be used as labels during learning stage, using Eq.~\ref{eq:solve} by considering only the coverage objective and without considering the temporal dimension. 
%The positions will be used as labels during the learning stage using Eq.~\ref{eq:solve} by considering only the coverage objective, without considering the temporal dimension.
%In fact, 
To this end, we utilized the UAV-BS placement approach proposed in~\cite{YalinizEY16}, where only $w_1$ is 1 (the other weights are zeroed).
%and only the first constraint in Eq.~~\ref{eq:solve} is necessary. 
However, without loss of generality, this scenario can be extended to include the other constituents of the generic reference model.

In Fig.~\ref{figure:CNN-Res}, UAV-BS locations given by our model and the optimization algorithm are presented for three test cases. Light blue disks and inverse triangles represent the initial and final locations of MUs, respectively. Dark blue dashes show the trajectory of the users. Red crosses and black diamonds indicate the locations obtained by the optimization algorithm and the proposed CNN approach, respectively. Yellow disks signify the proximity of the estimated locations. Fig.~\ref{figure:CNN-Res} shows that the UAV-BS locations estimated by the CNN-based approach are in close proximity of the optimal locations. 

\begin{figure}
\includegraphics[width = 3.5in]{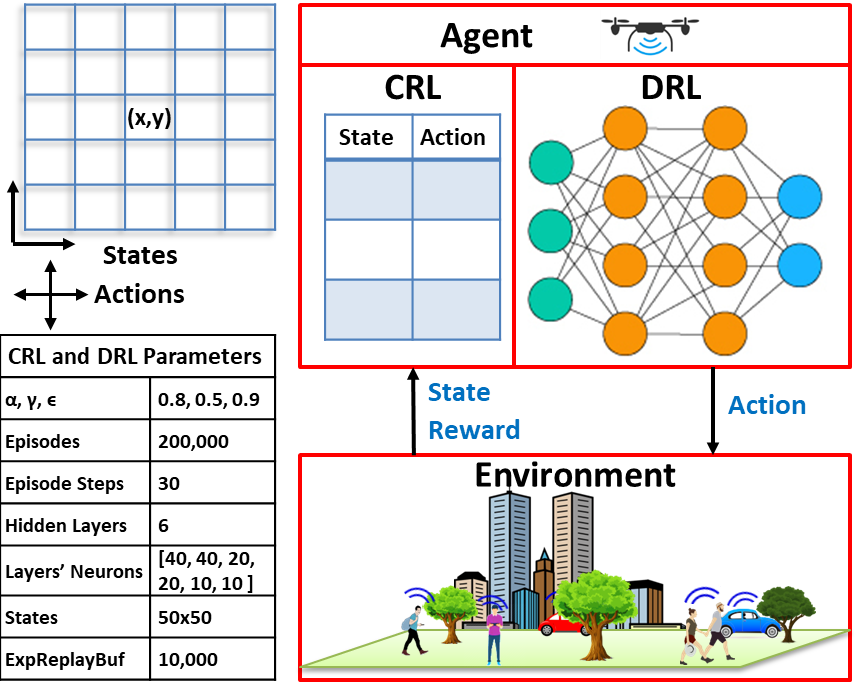}
\caption{RL approaches and their parameters.}
\label{figure:Rl}
\vspace{-0.3cm}
\end{figure}

\begin{figure*}[h]
    \vspace{-0.3cm}
    \hspace*{-2.4cm}
    \includegraphics[width = 8.8in]{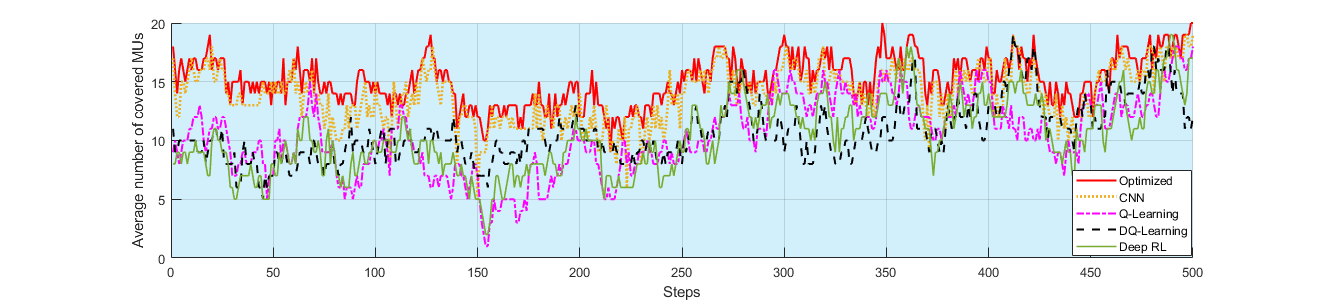}
    \caption{Coverage for 500 consecutive steps for a network with 30 MUs.}
    \label{figure:trajectory}
    \vspace{-0.3cm}
\end{figure*}

\begin{figure}[!h]
\vspace{-0.1cm}
\includegraphics[width = 3.5in]{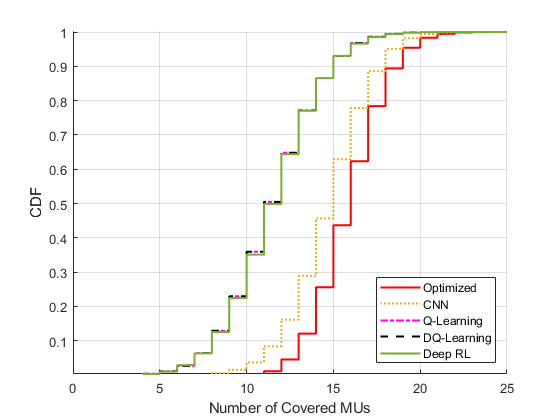}
\caption{CDFs of the covered MUs. }
\label{figure:cdf}
\vspace{-0.3cm}
\end{figure}

To investigate the performance of the CNN-based solution in comparison to RL-based approaches, we obtained results with DRL, Q-Learning, and Double Q-Learning~\cite{tang2021survey,chen2019artificial,ullah2020cognition}. Fig.~\ref{figure:Rl} illustrates the block diagram of the RL approaches, where CRL indicates the Conventional RL approaches (i.e., Q-Learning and Double Q-Learning). The parameters utilized to realize the RL approaches are also provided in Fig.~\ref{figure:Rl}. Furthermore, the discount factor for DRL is 0.99, and the loss function to update the weights (critical parameters) of the DNN is the mean square error. Fig.~\ref{figure:trajectory} presents the instantaneous coverage of different UAV-BS placement approaches for 500 consecutive steps, where each step is 4~s. The difference between the proposed approach and other learning methods can be as much as 10 covered MUs (e.g., at $50^{\textrm{th}}$ and $150^{\textrm{th}}$ steps). The Cumulative Distribution Function (CDF) of the covered MUs is illustrated in Fig.~\ref{figure:cdf} considering 18,000 test scenarios. The difference between the CNN approach and the optimal is 1 covered MU, on average, whereas, the average difference is, approximately, 4.5 covered MUs between RL-based approaches and the optimal. Although, our results reveal that the CNN-based approach outperforms other approaches for this particular use case, other learning approaches, especially DRL-based approaches, are shown to be highly successful in other use cases.
%\textbf{As a result, CNN has a promising success with respect to other approaches for this particular use case. Nevertheless, applying other learning approaches, especially DRL, may be more appropriate for some of the remaining use case scenarios.}

\section{Future Research Directions}
\label{section:research_challenges}
 
%In this study, we consider the positioning of a single UAV-BS. However, especially to cover large areas or to increase aggregate bandwidth offered to MUs, it is imperative to utilize multiple UAV-BSs. Furthermore, it is more efficient to position UAV-BSs in a coordinated fashion. Hence, placement of a plurality of UAV-BSs and integration of them to the SEN through learning approaches are promising research avenues. 
In this study, we consider the positioning of a single UAV-BS. However, especially to cover large areas or to increase aggregate bandwidth offered to MUs, it is imperative to utilize multiple UAV-BSs in a coordinated fashion. Hence, placement of a plurality of UAV-BSs and integration of them to SENs through learning approaches are promising research avenues.

%As mentioned above, 
Any UAV-BS placement approach should not be designed by assuming that the inputs are always consistent or error free. In fact, measurement and/or communication errors as well as man made intentional disinformation occur in real life deployments. 
Therefore, the training model should be 
%robust in presence of such imperfections. 
robust enough to handle such issues.
The proposed approach provides an effective solution to the robustness problem by averaging the locations of MUs. Yet, there are many other possible threats and errors. Nevertheless, improving the robustness and stability of a UAV-BS in the face of a wide range of threats/errors is an interesting research challenge.

%Many UAV-BS placement algorithms, in the literature, compute the locations of the UAV-BS considering the positions of MUs at a single time instant. During this computation, the users can move to other places. Thus, an autonomous UAV-BS should also predict the best locations considering the possible future positions of MUs. The proposed CNN model can be used for this purpose. Considering the temporal dimension of the UAV-BS positioning problem, estimating future locations of MUs, accurately, would be invaluable for making efficient trajectory estimation for a UAV-BS. In fact, people, generally, do not move completely randomly, hence, it is possible to predict the future locations of MUs reasonably well. 

ML-based UAV-BS positioning algorithms necessitate the use of training data sets. However, there are no publicly available datasets for this purpose. While it is possible to generate synthetic datasets (as in this study) or to create datasets based on extensive measurement 
%campaigns. Both approaches 
campaigns, both of these approaches have advantages as well as disadvantages. Synthetic datasets are easy to generate, yet they can fail to represent the real world conditions precisely. Measurement-based datasets reflect actual scenarios more accurately, however, they are both time consuming and inflexible (i.e., they cannot be modified, easily, to model conditions other than the one they are actually obtained from). Nevertheless, it is extremely important to create a wide range of realistic datasets to be used in the analysis of NGWN scenarios with UAV-BSs, preferably, endorsed by an international communications organization such as IEEE or ITU.

Although DNNs require much less computation during the inference stage when compared to the training stage, it is still challenging to use them on resource constrained embedded systems (such as the ones in UAV-BSs) due to their memory requirements. Therefore, it is of utmost importance to reduce the resource requirements of DL-based approaches to be used by UAV-BSs without sacrificing their performances, significantly.

%\textbf{The location where the learning is carried out is an interesting problem to consider. Since UAV-BS's have limited memory and processing capabilities, they can not converge during training. However, using collaborative learning techniques like Federated Learning at Edges mitigates the challenges of training. Another option is to use cloud which provides the fastest convergence but communication delay is an issue.}   

The point where the learning is carried out is an interesting problem to consider. UAV-BSs have limited memory and processing capabilities, therefore, it is inefficient to employ them for training. However, using collaborative learning techniques such as federated learning at edges could mitigate the challenges of training.

\section{Conclusion}
\label{section:conclusion}

Positioning a UAV-BS for the optimal operation of the network is a challenging problem. In this study, we first introduced a comprehensive optimization model for UAV-BS positioning. Next, we presented a CNN-based approach for the efficient solution of a special case of the model considering its coverage aspect. %Performance evaluations of the proposed CNN-based approach in comparison to RL-based approaches reveal 
We then evaluated the performance of the proposed CNN-based approach against RL-based approaches, which revealed that the CNN-based approach leads to the optimal solutions with negligibly low differences. Yet, the CNN-based approach is four orders of magnitude faster than the optimization approach. %\textbf{The complexity of the proposed approach only depends linearly on grid size, O(N), whereas, the complexity of an MINLP problem is NP.} 
The complexity of our proposed approach only depends linearly on grid size.
Nevertheless, there are still many challenges to be addressed for fulfilling the promise of autonomous UAV-BSs, which we also outlined in this study. 

\bibliographystyle{IEEEtran}
\bibliography{UAV-DL-ComMag}

\footnotesize
\vspace{0.3cm}
\noindent\textbf{A. Murat Demirtas} (ademirtas@etu.edu.tr) is an assistant professor in the Department of Electrical and Electronics
Engineering, TOBB University of Economics and Technology, Ankara, Turkey. His current research areas are machine learning for communications, wireless networks, wireless communications, optimization, video coding, video streaming and rate-distortion theory.

\vspace{0.3cm}
\noindent\textbf{M. Saygin Seyfioglu} (msaygin@cs.washington.edu)
received his B.S. degree in electrical and electronics engineering and M.S degree in electrical engineering from TOBB University of Technology and Economics, Ankara, Turkey, in 2015 and 2017, respectively. He is a recipient of Fulbright PhD Fellowship and currently pursuing his PhD at University of Washington. His research interests include signal processing, deep learning and computational biology. 

\vspace{0.3cm}
\noindent\textbf{Irem Bor-Yaliniz} 
(irembor@sce.carleton.ca) received her B.Sc. and M.Sc. degrees in electrical and electronics engineering from Bilkent University, Turkey, in 2009 and 2012, respectively, and her Ph.D. from Carleton University, Canada in 2020. She has also been with Huawei Ottawa Research Development Centre since 2017. She is the co-inventor of 15+ patent
applications worldwide, and received scholarships through the Natural
Sciences and Engineering Research Council of Canada (NSERC) and the Queen
Elizabeth II Scholarship in Science and Technology. She became the first
Canadian to be chosen as a Rising Star in Computer Networking and
Communications by N2Women in 2019.

\vspace{0.3cm}
\noindent\textbf{Bulent Tavli} (btavli@etu.edu.tr) is a professor in the department of electrical and electronics engineering, TOBB University of Econonmics and Technology, Ankara, Turkey. His current research areas are network science, telecommunications, smart grid, security and privacy, machine learning, optimization, algorithms, and blockchain.

\vspace{0.3cm}
\noindent\textbf{Halim Yanikomeroglu} (halim@sce.carleton.ca) is a full professor in the Department of Systems and
Computer Engineering at Carleton University, Ottawa, Canada.
His research interests cover many aspects of 5G/5G+ wireless networks, including layers 1-3, and the architecture.
His collaborative research with industry has resulted in 37 granted patents.
He is a Fellow of the IEEE, a Distinguished Lecturer for the IEEE Communications Society, and a 
Distinguished Speaker for the IEEE Communications Society and the IEEE 
Vehicular Technology Society.

\end{document}